\documentclass[preprint,showpacs,showkeys,preprintnumbers,amsmath,amssymb]{revtex4}
\usepackage{graphicx}% Include figure files
\usepackage{dcolumn}% Align table columns on decimal point
\usepackage{bm}% bold math
\usepackage{float}
\begin{document}

%\preprint{Version 0.7}

\title{Morphology of Superconducting FeSe thin films deposited by co-sputtering and MBE}% Force line breaks with \\

\author{Eike Venzmer}
\author{Alexander Kronenberg}
\author{Janek Maletz}
\author{Martin Jourdan}
 \email{Jourdan@uni-mainz.de}
\affiliation{Institut f\"ur Physik, Johannes Gutenberg-Universit\"at, Staudinger Weg 7, 55128 Mainz, Germany}

\date{\today}% It is always \today, today,

\begin{abstract}
The presumably unconventional superconductor $\beta$-FeSe was deposited by radio frequency sputtering and molecular beam epitaxy (MBE) from two elementary sources. Superconducting thin films were grown in (001)-orientation on MgO(100) and YAlO$_3$(010) substrates. The morphology of the samples was studied and directly related to the superconducting properties of the $\beta$-FeSe thin films. The MBE grown thin films show microcracks down to the substrate and a roughness of $\simeq$ 100~nm. In contrast, sputter deposited superconducting thin films show a smooth surface with almost no precipitates. In both cases resistive superconducting transitions with critical temperature up to $\text{T}^\text{midpoint}_\text{c} \simeq 8.5\;\text{K}$ were observed. However, the smoothness of the sputter deposited films is crucial for future surface dependent investigations.
\end{abstract}

%Use showkeys class option if keyword
\maketitle

\section{Introduction}
%short general introduction to Ironpnictides and thin films:
Short after the discovery of the first superconducting iron-pnictide, the compound LaFeAsO$_{1-\text{x}}$F$_\text{x}$ \cite{Kam08}, the structurally most simple representative of this class of materials, the compound $\beta$-FeSe, was identified \cite{Hsu08}. Soon the first thin film samples were grown, which, additional to potential technical applications, are important for investigations of the presumably unconventional mechanism of superconductivity. However, there are still many limitations of the thin film quality, including homogeneity and stoichiometry, as e.\,g.\ described in \cite{Hai14}. Please note that in many cases the usefulness of epitaxial thin film samples depends on their morphology. This is the aspect on which this paper is focused.\\  
Up to now epitaxial $\beta$-FeSe thin films were deposited by several techniques: Low-pressure metal organic chemical vapor deposition (LP-MOCVD) was used by Liu \textit{et al.}\,\cite{Liu07} and Wu \textit{et al.}\,\cite{Wu08} for growing FeSe on GaAs(001) and SiO$_2$, respectively. Wu et al. presented a strong dependence of the morphology on the substrate temperature during deposition. However, no information on the superconducting properties of these thin films was shown. Tkachenko \textit{et al.} obtained very rough superconducting thin films of $\beta$-FeSe on LaAlO$_3$ and SrTiO$_3$ by the use of a high gas pressure trap system (HGPTS) \cite{Tka09}. Pulsed laser deposition (PLD) is an often used method for the deposition of $\beta$-FeSe on a broad range of substrates resulting in samples with transition temperatures up to $\text{T}_\text{c}^\text{onset}=11.8\;$K \cite{Nie09, Han09, Wan09, Wu09, Yua13}. Information about the morphology of PLD grown thin films was published by Han \textit{et al.}\ \cite{Han09} only, who also presented  strong dependence on the substrate temperature and generally a rough surface associated with good superconducting properties. A selenization technique was utilized by Takemura \textit{et al.}\ for the preparation of non-superconducting thin films of FeSe on GaAs \cite{Tak97}. Qing \textit{et al.}\ fabricated superconducting $\beta$-FeSe thin films on LaAlO$_3$ \cite{Qin11}. Again their surfaces showed a rough morphology with typical grain sizes in the magnitude of $\simeq1\mu$m. Schneider \textit{et al.}  presented thin films deposited by radio-frequency (rf) sputtering from one stoichiometric FeSe-target and an additional Se-target for tuning the Se content resulting in superconducting samples with $\text{T}_\text{c} = 3-11\;\text{K}$ strongly depending on film thickness \cite{Sch13}. Scanning electron microscopy (SEM) images of their films show large precipitates with grains in size within the range of $0.6-1.4\;\mu\text{m}$.\\ 

\section{Preparation and structural characterization}
We prepared epitaxial $\beta$-FeSe thin films with varying composition by molecular beam epitaxy (MBE) and radio frequency~(rf) sputtering resulting in similar superconducting, but very different morphological properties.\\
Within the MBE growth process, Fe from an electron-beam evaporator and Se from an effusion cell were co-evaporated. With a growth rate of about $0.1\;$nm/s the typical sample thickness amounted to 500\,nm. Good results regarding crystal growth and superconductivity were obtained using YAlO$_3$(110) substrates, no superconducting samples were obtained on MgO(100) substrates. The substrate temperature during deposition amounted to $\text{T}_\text{substrate}=350\,^\circ$C. More details on the MBE growth process, structural characterization and magnetotransport measurements can be found in \cite{Jou10}.\\
Alternatively, $\beta$-FeSe thin films were rf-sputtered from two elementary 2\,inch targets (Fe: Mateck GmbH, purity $99.99\%$; Se: Lesker, purity $99.999\%$) using AJA International A320-XP-MM sputtering sources. Best results were obtained with rf-powers of 7-8\,W for selenium and 13-14\,W for iron. Higher rf-powers of the Se source result in partial melting of the target. With an Ar sputter gas pressure of 0.0073\,mbar a growth rate of 0.006\,nm/s was obtained. Typically, the samples were prepared with thicknesses of 20-100\,nm. The optimized substrate temperature amounts to $\text{T}_\text{substrate}\approx 350\,^\circ$C, which is the same value as for MBE grown thin films. In contrast to MBE growth, epitaxial growth of superconducting thin films was not only obtained on YAlO$_3$(010) substrates, but also on MgO (100) despite a huge lattice mismatch of 11.8\% of this substrate material.\\ 
By depositing from two sources, it is possible to generate a gradient of the Fe/Se-content ratio across one or several substrates during a single deposition process. This gradient is much stronger in the sputter deposited than in the MBE grown thin films. Measurements by energy dispersive X-ray spectroscopy (EDX) indicate differences of $\approx 20\%$ in the Fe/Se-content ratio for co-sputtered FeSe along a distance of 10\,mm. This allows the study of stoichiometry dependent properties on a single thin film sample. Due to the relatively thin film thickness compared to the information depth of EDX no absolute values for the sample stoichiometry could be determined.\\ 
Structural investigations by X-ray diffraction (XRD) of MBE grown samples demonstrated the epitaxial growth of $\beta$-FeSe in (001)-direction \cite{Jou10}.
The co-sputtered samples grow in the same (001)-direction as shown in Fig.\,1.
The 20\% gradient in the Fe/Se-content ratio is reflected in the XRD $\theta /2\theta$-scans, if the samples are partially covered by a Cu mask during the scans. Whereas for the scan displayed in Fig.\,1\,a, red line the complete sample was illuminated by the x-rays, only part of the thin film (width $\simeq 2.5$\,mm) contributed to the scan displayed in Fig.\,1\,a, black line. It is obvious that within this selected sample region the impurity phase of Fe$_3$Se$_4$ (marked * in Fig.\,1\,a) vanished completely. The impurity phase of $\text{Fe}_7\text{Se}_8$ still appears but with decreased intensity. The inset of Fig.\,1\,a shows an extremely narrow rocking curve ($\omega$-scan) of the FeSe(001) peak with $\rm{\Delta \omega \simeq 0.02^{\circ}}$. From this a lower bound for the structural in-plane correlation length of $2\pi / \Delta q_{\omega} = c/(2\tan(\Delta \omega /2))\simeq 1.5{\rm \mu m}$ can be estimated.\\
The in-plane order and epitaxial relation of the FeSe(001) thin films on MgO(100) is obvious from the XRD $\phi$-scan of the off-specular FeSe(101) peaks, which are found at the same $\phi$-values as the off-specular (220) peaks of the substrate (Fig. 1\,b). Thus the in-plane FeSe(100) and MgO(010) directions are aligned parallel. Weak additional FeSe(101) peaks are visible at $45^\circ$ relative in-plane rotation, corresponding to in-plane rotated grains of FeSe. The lattice constants can be calculated from the positions of the FeSe(00l) and the FeSe(101) peaks as $a=3.765(10)$\r{A} and $c=5.53(1)$\r{A}. The lattice parameter $a$ is in agreement with that presented by Hsu \textit{et al.} for bulk material, while the c-axis is slightly longer than the bulk value of c=5.4847(1)\r{A}) \cite{Hsu08}.\\ 
\begin{figure}[H]
	\begin{center}
	
			\includegraphics[width=0.5\columnwidth, angle=0]{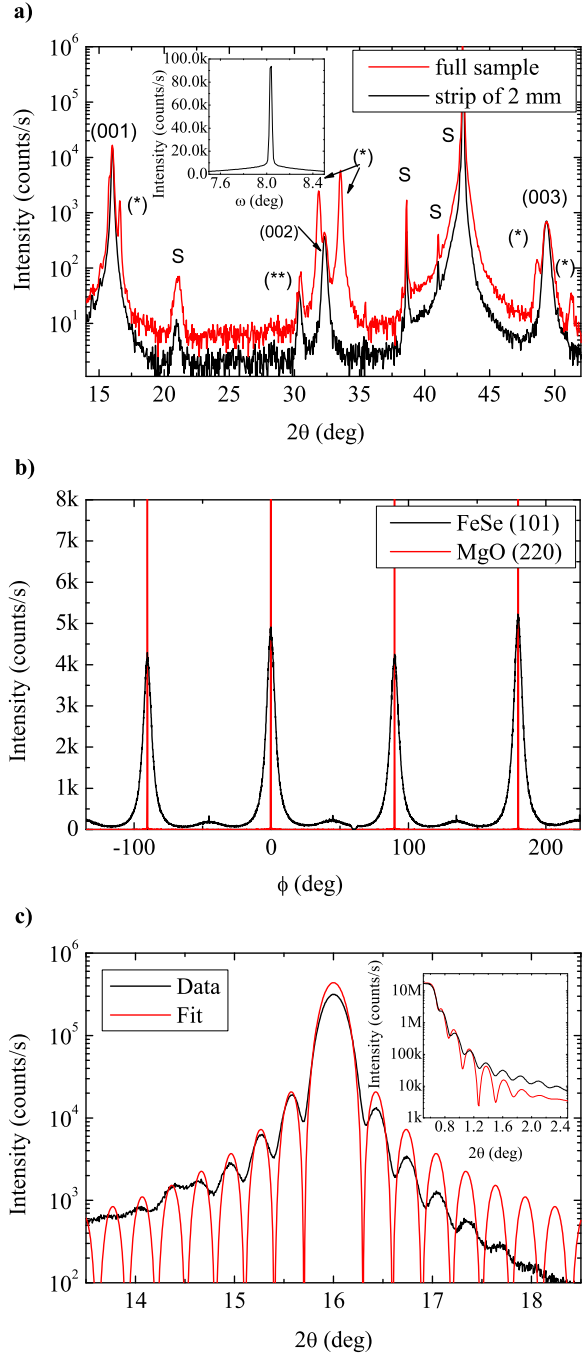}
				
			\caption{a) X-ray $\Theta/2\Theta$-scan of a FeSe thin film grown by co-sputter deposition on a MgO(100) substrate of one full sample and a strip of $2\,\text{mm}$ only. The (001)-axis of FeSe is aligned perpendicular to the substrate surface. Additionally to the peaks produced by the substrate, impurity phases of Fe$_3$Se$_4$ (*) and Fe$_7$Se$_8$ (**) were identified. The inset shows a rocking curve of the FeSe(001) peak. b) $\phi$-scan of the off-specular (101) peak of the same thin film as in a) (black line). Additionally a $\phi$-scan of the off-specular (220) peak of MgO is shown (red line). c) X-ray $\Theta/2\Theta$-scan of a FeSe thin film deposited by co-sputtering on a MgO(100) substrate. The (001)-peak shows pronounced Laue oscillations demonstrating coherent crystallographic order. The inset shows an X-ray reflectometry measurement and simulation of the same sample.}
	\end{center}
\end{figure}	
The film thickness was determined by the evaluation of XRD Laue oscillation patterns and by X-ray reflectometry (XRR) (Fig.\,1\,c). From the fitting of the Laue oscillations a thickness of 29.9\,nm was obtained, whereas fitting the XRR pattern resulted in a thickness of 31.5\,nm. The good agreement of both values is evidence for coherent crystallographic order over the complete geometrical thickness of the thin film.

\section{Morphology and Superconductivity}
The morphology of the rf-sputtered FeSe thin films will be discussed in comparison with MBE grown samples. Independent from the deposition method, the FeSe thin film morphology depends strongly on the sample stoichiometry. Fig.\,2\,a shows a scanning electron microscopy (SEM) image of an MBE grown sample with a nearly optimum stoichiometry resulting in a high $\text{T}_\text{c}^\text{midpoint}\simeq 8\;$K. This film has a flaked morphology with some large segregations and is interspersed with an orthogonal pattern of micro-cracks. Atomic force microscopy (AFM) (Fig.\,3, left) provides evidence that these cracks do not fully separate the different thin film sections but nevertheless partially go down to the substrate surface. Additionally, the AFM also shows the enormous roughness of the FeSe thin film with local thickness variations up to half of the total film thickness.\\
However, sputter deposited FeSe thin films with optimized stoichiometry, again reflected in a high $\text{T}_\text{c}^\text{midpoint}\simeq 7.5\;$K, have a much smoother surface as shown in Fig.\,2\,c (SEM) and Fig.\,3, right, (AFM). Most importantly, the micro-crack pattern of the MBE samples is absent in theses films. Thus rf-sputter deposition results in a clearly improved thin film morphology, not only compared to our own MBE grown samples but also compared to all other FeSe thin films mentioned in the introduction. This is probably related to the relatively high kinetic energy of the adatoms in rf-sputtering compared to all other thin film deposition methods. 
The morphology depends strongly on the stoichiometry. On the same sputter deposited sample, but at different positions corresponding to Fe/Se ratios which are reduced (Fig.\,2\,b) or increased (Fig.\,2\,d) by $\simeq 5\%$ compared to Fig.\,2\,c, completely different morphologies were obtained. As already discussed above, small deviations of the stoichiomety result in the formation of impurity phases, which possibly form the segregations visible in the SEM images.\\
Due to the gradient in the Fe/Se ratio across the substrate also the transport properties including the superconducting transition temperature T$_c$ vary across the thin films. The resistivity of the thin film discussed above concerning its morphology was measured by attaching wires to the four corners of the $10 \times 10$\,mm sample and by passing a current parallel and perpendicular to the Fe/Se gradient (Fig.\,4a, inset). For both directions similar temperature dependencies of the resistivity R(T) were obtained (Fig.\,4a). However, whereas for the current direction perpendicular to the gradient a full superconducting transition with $\text{T}^{\text{midpoint}}_{\text{c}}=7.5\;\text{K}$ and a relative narrow width of $\Delta\text{T}_\text{c} \simeq 1.1\;\text{K}$ was obtained, for the parallel direction only a partial transition was measured. This is in agreement with the assumption of a superconducting section of the sample, which forms a strip aligned perpendicular to the Fe/Se gradient. Photolithographic patterning of the sample was used to generate well defined regions for resistivity measurements of the thin film. The three lithographically defined strips of $0.1\,\mu\text{m}$ width (inset of Fig.\,4\,b) show different R(T) curves with strip 3 corresponding to Fe excess showing an increase in resistivity at low temperatures and no superconductivity. Strips 1 and 2 both become superconductive with the maximum $\text{T}^{\text{midpoint}}_{\text{c}}=5.8\;\text{K}$ corresponding to the central strip 2. This is a reduced value compared to the non-patterned sample which is either due to aging of the thin film during lithography or the section with the highest T$_c$ was missed by all patterned strips.\\
The critical current density for strip 2 is measured to be $\text{j}_\text{c}\approx 6.8\cdot 10^3 \text{A}/\text{cm}^2$ at a temperature of $\text{T}=4.2\;\text{K}$ which is in the same range as measured by Lei \textit{et al.} in $\beta$-FeSe single crystals with $\text{j}_\text{C}\approx 10^4 \text{A}/\text{cm}^2$ at $\text{T}=1.8\;\text{K}$ \cite{Lei11} thus indicating a substantial superconducting volume in the thin films.\\
\begin{figure}[H]
	\begin{center}
	
		\includegraphics[width=1.0\columnwidth, angle=0]{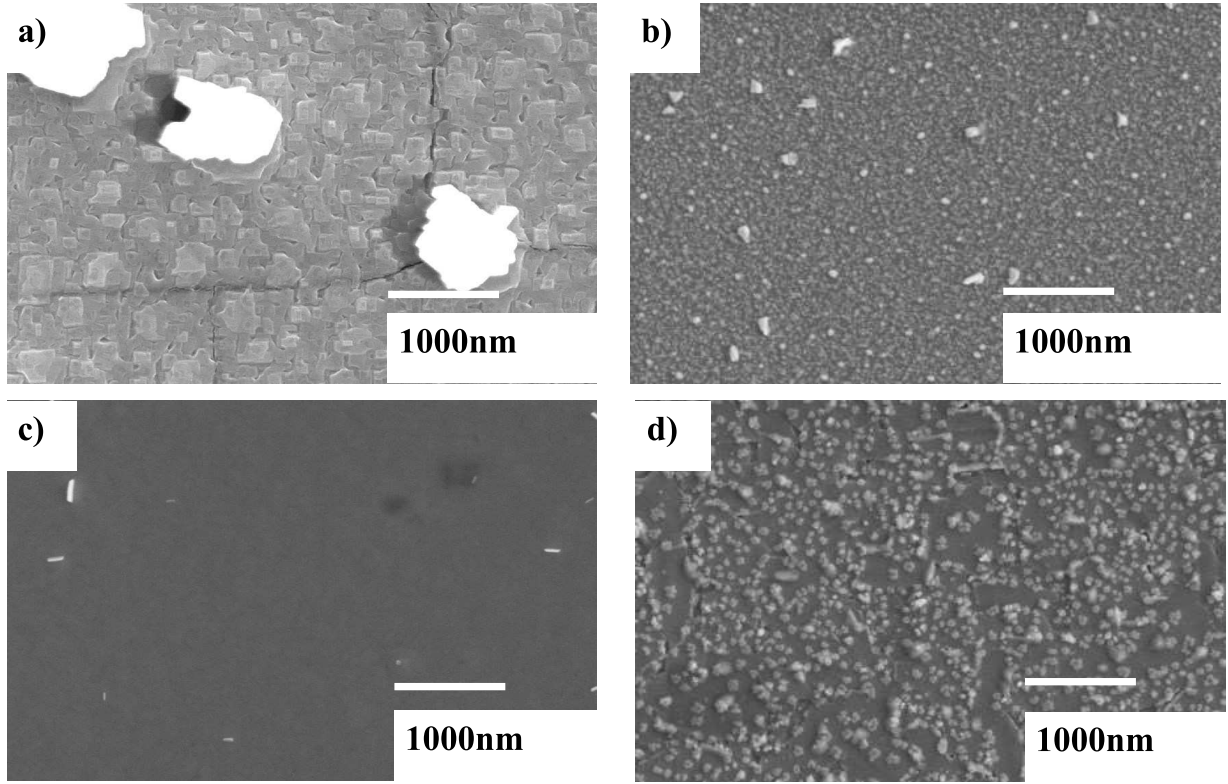}
		
		\caption{SEM images of FeSe thin films prepared by MBE (\textit{a}) and sputter deposition (\textit{b, c and d}). SEM images of co-sputtered thin films are taken at different positions along a stoichiometric gradient. The Fe/Se ratio decreases by $\simeq${15\;\%} from \textit{b} to \textit{d}.}
	\end{center}
\end{figure}
	
\begin{figure}[H]
	\begin{center}
	
				\includegraphics[width=0.9\columnwidth, angle=0]{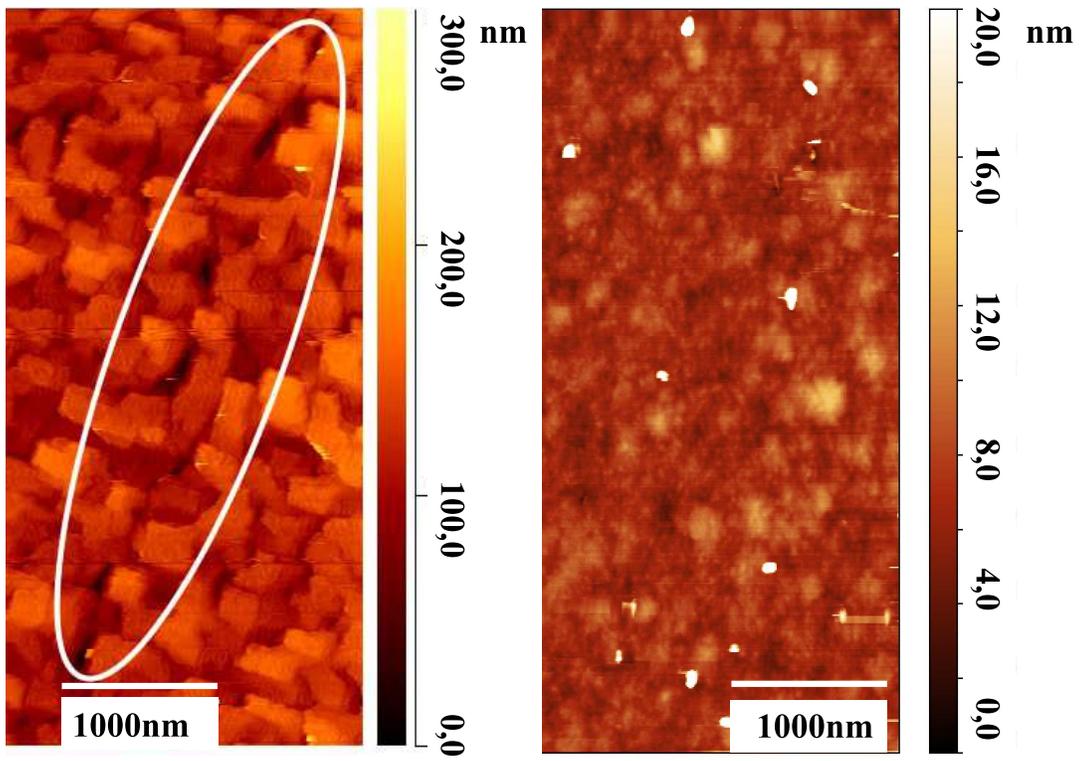}
				
				\caption{\textit{Left:} AFM image of a FeSe thin film prepared by MBE. \textit{Right:} AFM image of a FeSe thin film deposited by rf-sputtering.}
	\end{center}
\end{figure}

\begin{figure}[H]
	\begin{center}
	
		\includegraphics[width=0.8\columnwidth, angle=0]{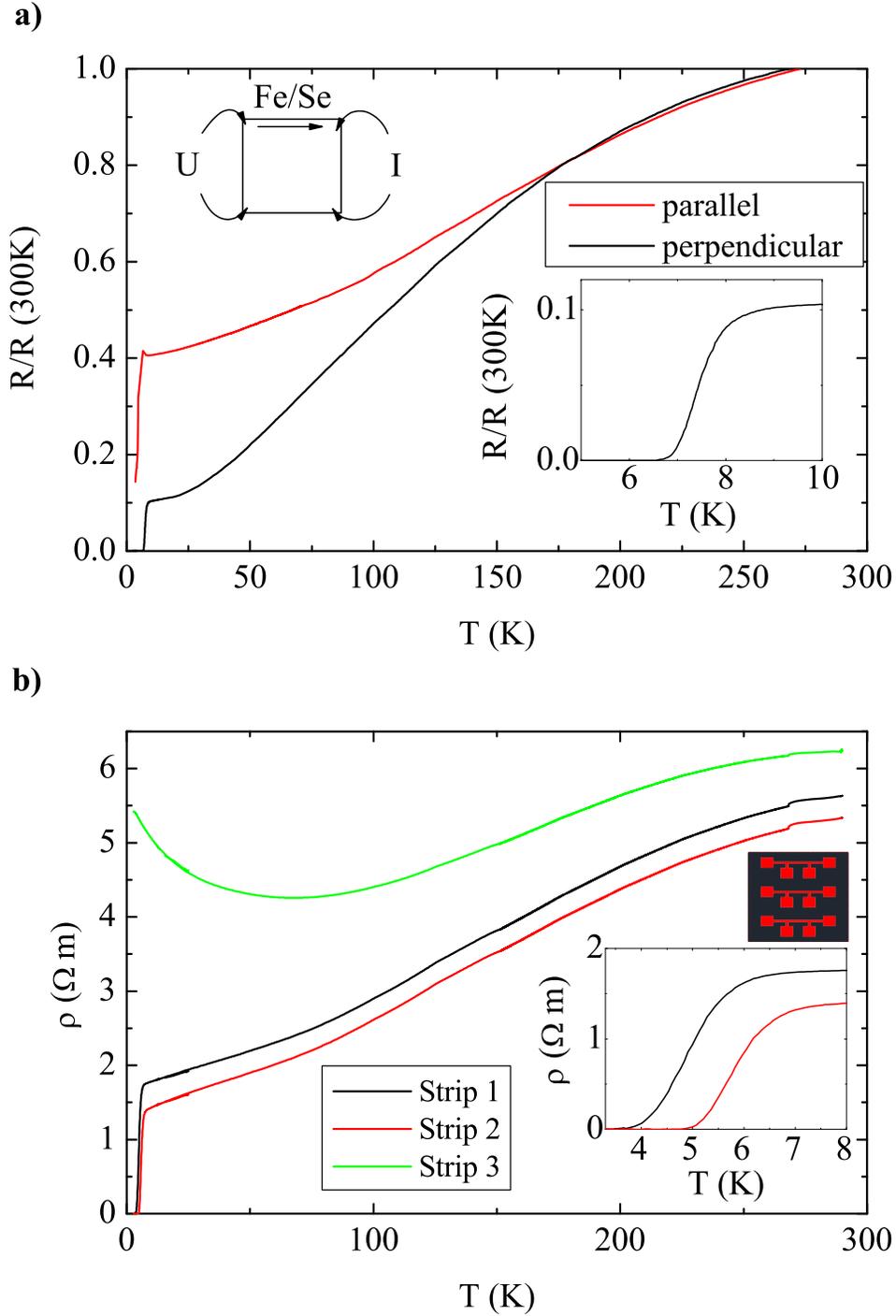}

		\caption{a) Normalized resistivity of a FeSe thin film deposited by rf-sputtering, either with the current in the direction of the stoichiometric gradient (red line) or perpendicular to it (black line, configuration shown in the upper inset). The lower inset shows the superconducting transition for the perpendicular current direction at about $\text{T}_c=8.5\text{K}$. b) Specific resistance for FeSe strips at different positions along the stoichiometric gradient patterned by photolithography. The inset shows the superconducting transitions for strip 1 (black line) and strip 2 (red line).}
	\end{center}
\end{figure}

\section{Conclusion}
Epitaxial superconducting FeSe thin films were deposited by rf-sputtering from two standard sputtering sources. Compared to MBE-grown and laser ablated thin films the sputtered samples show a much smoother and micro-crack free morphology. The morphology as well as the transport properties including the superconducting transition temperature depend strongly on the stoichiometry. Provided that the correct Fe/Se ratio is adjusted, rf-sputter co-deposition results in a promising morphology of FeSe(001) thin films, which allows new surface or interface based investigations of the unconventional superconducting state.\\
 
Financial support by the DFG (Jo404/6-1) is acknowledged.

\end{document}